# SLA-aware Interactive Workflow Assistant for HPC Parameter Sweeping Experiments

Bruno Silva, Marco A. S. Netto, Renato L. F. Cunha
IBM Research

## ABSTRACT

A common workflow in science and engineering is to (i) setup and deploy large experiments with tasks comprising an application and multiple parameter values; (ii) generate intermediate results; (iii) analyze them; and (iv) reprioritize the tasks. These steps are repeated until the desired goal is achieved, which can be the evaluation/simulation of complex systems or model calibration. Due to time and cost constraints, sweeping all possible parameter values of the user application is not always feasible. Experimental Design techniques can help users reorganize submission-execution-analysis workflows to bring a solution in a more timely manner. This paper introduces a novel tool that leverages users' feedback on analyzing intermediate results of parameter sweeping experiments to advise them about their strategies on parameter selections tied to their SLA constraints. We evaluated our tool with three applications of distinct domains and search space shapes. Our main finding is that users with submission-execution-analysis workflows can benefit from their interaction with intermediate results and adapt themselves according to their domain expertise and SLA constraints.

## 1. INTRODUCTION

Evaluation of test scenarios and model calibration are common in several industries including finance, aerospace, health, and energy. Normally users run applications that contain a set of parameters, each able to assume a broad range of values. These applications are known as parameter sweeping applications or parametric applications. In this context, it is necessary to run several application instances varying their parameter values to find a combination that meets a given specification or optimization criteria.

A popular practice in this context is to employ Design of Experiments (DOE) techniques [21] to select parameter values to be evaluated. These methods help users understand the impact of each parameter value on the generated output. As users can check the results produced by each execution, they can utilize the intermediate results to select the next parameter values to be evaluated. The knowledge on the impact of parameter values helps users explore the search space more effectively. Nevertheless, to the best of our knowledge, existing work does not exploit human knowledge/feedback and DOE techniques to assist users on parameter selection decisions under these submission-execution-analysis workflows with Service Level Agreement (SLA) constraints.

Over the years, optimization methods have been created to solve complex problems in industry and academia. However, fully automatic solutions face some obstacles to obtain satisfactory results due to the following reasons [20]:

- It is hard to obtain/create an optimization model that reflects all aspects of a real-world problem, especially when multi-criteria objectives are involved;
- Even when the optimization model is adequate, the time/cost to find an optimal solution may violate user's time or cost restrictions;
- The analyst expertise and creativity are hard to code. Then, for problems that involve complex and important decisions (e.g., financial trading decisions), even fully automatic solutions must be, in the end, validated by humans to be adopted.

Generally, it is prohibitive to run application instances with all possible parameter values due to time and cost constraints—cost, in particular, becomes a key factor considering execution of these applications in outsourced environments (e.g., public clouds). Additionally, optimization techniques can be adopted to suggest parameter values in order to reduce the number of evaluated scenarios and consequently reduce the experiment costs. Therefore, SLA-aware prediction methods and mechanisms to suggest parameter values would help engineers and scientists to evaluate parametric applications.

In this work, interactive optimization methods are proposed to take advantage of human expertise/creativity and processing power to solve optimization problems. We study how the decision maker's experience can be leveraged to find solutions for parameter sweeping applications. The analyst's decisions can be impacted by the DOE analysis that shows which parameters have more impact on the output. For instance, an experiment explorer tool can show which parameter values cause more impact on experiment results and the user can select suitable parameter value combinations based on this information.

We introduce a tool, called Copper (**C**ognitive **Op**timizer and **P**arameter Explor**er**), to help users in the evaluation of search spaces considering human feedback and domain expertise. As new samples are evaluated, Copper updates the impact of each parameter value on the results and generates new samples for evaluation by using derivative free optimization (DFO) methods [4]. Moreover, clustering techniques are employed to evaluate the regions of search space in which the user is interested and to propose the user to change her approach when the SLA is predicted to be violated. The main contributions of this paper are therefore:

- A tool to assist users in the evaluation of scenarios with parameter sweep applications using human feedback and domain expertise (§ 3);
- Clustering methods to evaluate user behavior and propose strategy changes in case of imminent SLA violations (§ 3.2);
- Evaluation of the tool using three applications from different domains and search space shapes (§ 4).

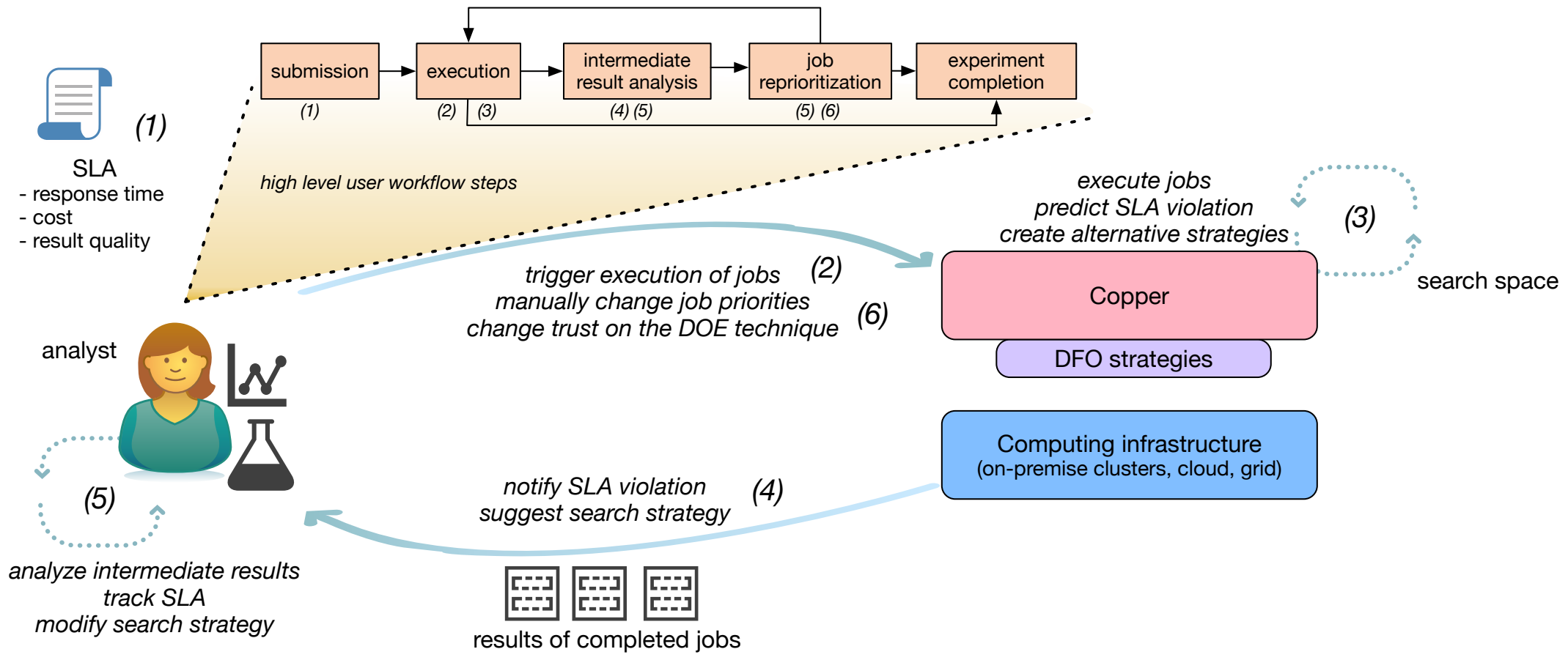

**Figure 1: User workflow with job submission, job execution, and intermediate result analysis steps.**

## 2. BACKGROUND

This section presents research efforts related to our work and an overview of the problem investigated in this paper.

### 2.1 Related Work

Assisting users in executing applications has been a research goal of several groups. Related to our work there are efforts from multiple areas: computational steering, human-guided search, workflow management, design of experiment, interactive optimization, among others.

Computational steering [5, 11–13, 22, 32] aims at providing users with tools that enable parameter reconfiguration while experiments are in progress. Parker and Johnson [26] introduced a system called SCIRun that uses a dataflow programming model and visual programming to simplify the tasks of creating, debugging, optimizing, and controlling complex scientific simulations. Van Wijk *et al.* [30] highlighted that the implementation of computational steering in practice is hard. To overcome this problem they implemented an environment in which a data manager entity facilitates the interaction between the application and the steering components. Chin *et al.* [2] incorporated computational steering in mesoscale lattice Boltzmann simulations and showed the benefits of their work. They discussed that large scale simulations require not only computational resources but tools to manage these simulations and their produced results, what they called *simulation-analysis loop*. Netto *et al.* [23] introduced a scheduler system able to automatically offer more resources to parametric application jobs based on the quality of their intermediate generated results so as users could get faster to their desired goal. More recently, Mattoso *et al.* [18] surveyed the use of steering in the context of High Performance Computing (HPC) scientific workflows highlighting a tighter integration between the user and the underlying workflow execution system. Another area to help users is the development of workflow management systems [6, 17].

An important research topic related to our work is optimization assisted by humans [19, 20]. Meignan *et al.* [20] provided a detailed survey and taxonomy of efforts in interactive optimization applied to operations research. They explored the different roles a user can have in an optimization process, such as adjusting or adding new constraints and objective, helping on the optimization process itself, and guiding the optimization process by providing information related to decision variables. Meignan and Knust [19] proposed a system that employs analyst feedback on the optimization to use as long-term preferences for future runs. Nascimento and Eades [7] proposed a framework for humans to assist the optimization process via inserting domain knowledge, escaping from local minimum, reducing the search space to be explored, and avoiding ambiguity for optimal multi-solutions. Researchers have also explored visualization techniques to add humans in the optimization process [3, 9, 15, 29]. WorkWays [24, 25] is a science gateway with human-in-the-loop support for running and managing scientific workflows.

Several efforts in the design of experiments happened over the last years. Kleijnen *et al.* [16] developed a survey and a user guide on advances in this area until 2005. Using fractional factorial design technique, Abramson *et al.* [1] developed a system to facilitate parameter exploration and support for abstracting the underlying computing platform.

Our research is based on existing work of these aforementioned efforts in order to build a tool able to provide feedback to the user on how her interaction with the parameter sweeping experiments impacts the defined time (and cost) constraints.

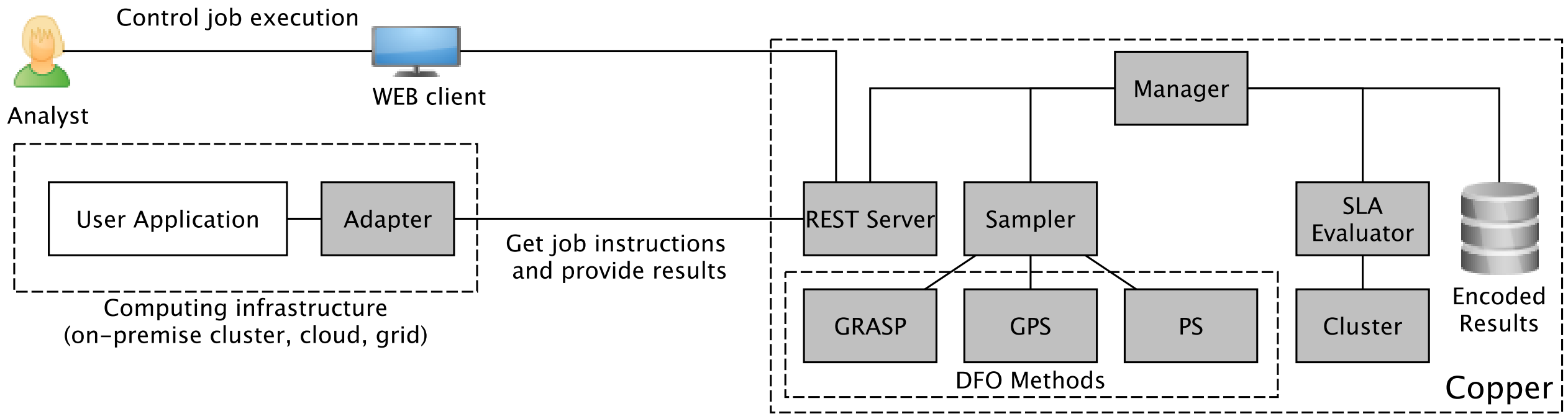

**Figure 2: Overview of Copper and application adapter.**

## 2.2 Problem Description

Problems from several industries are tackled by users, also named *analysts*, running complex computer simulations and optimizers that constitute a *software* with a set of *parameters*, where each parameter can receive different *values*. We consider applications with $n$ parameters and each parameter has a finite discrete domain $\mathbb{D}_i$ where $i \in \{1, \cdots, n\}$. A function $f : \mathbb{D}_1 \times \cdots \times \mathbb{D}_n \rightarrow \mathbb{R}$ is adopted to evaluate the quality of parameter values and is specific for each parametric application. The considered optimization problem $P$ is described as follows:

$$P \left| \begin{array}{l} max\ f(x) \\ x \in \mathbb{D}_1 \times \cdots \times \mathbb{D}_n \end{array} \right.$$

We assume the user is able to generate $f(x)$ according to the simulation output. For instance, if the output $g(x)$ of a simulation process is a real value that should be minimized, then $f(x)$ can be assigned as $f(x) = -g(x)$ and the problem definition remains the same. If the parametric application $a$ generates multiple outputs (e.g., $a : \mathbb{D}_1 \times \cdots \times \mathbb{D}_n \rightarrow \mathbb{O}_1 \times \cdots \times \mathbb{O}_m$), a wrapper function $h : \mathbb{O}_1 \times \cdots \times \mathbb{O}_m \rightarrow \mathbb{R}$ should be employed to represent the behavior of $f$ (i.e., $f(x) = h(a(x))$).

For instance, suppose the analyst wants to calibrate a simulation model that generates a list of predicted values over a given period. In this case, the output of the parametric application corresponds to a list of values, one for each time instant. A wrapper function (e.g., Root Mean Squared Error) can be employed to generate a single quality value to compare the difference between the generated values and a reference series.

Running all possible values for each parameter can be unfeasible—even with large high performance computing machines. Therefore, users need strategies to select a subset of values for each parameter that covers parts of the exploration space that will give them enough information to answer their questions. Such questions can be: what is the best set of parameter values that provides a (close-to-)optimal solution or what are the parameter relationships that have more influence on the phenomenon being analyzed. The strategies for parameter-value selections can be defined by Design of Experiments (DOE), which corresponds to a systematic method to determine the relationship between parameters (also known as *features*) affecting a process and the output of that process.

Analysts may have initial insights about the regions of search space that lead to good results. However, these insights may be wrong, and can change as long as intermediate results are evaluated. For several problems, the codification of human insights is not feasible due to the following reasons. These insights come from observations of historical data which may not be available. Even when this data is available, the time to analyze and codify the human insight may not be affordable. Human insights can be based on domain expertise, and the analyst has no experience/time to explain or codify this knowledge. Therefore, a mixed strategy that combines human expertise and optimization approaches may be useful to help analysts in the evaluation of complex optimization problems.

In this paper, we introduce a tool and a set of techniques to help scientists and engineers to properly prioritize their experiments, which can comprise traditional HPC jobs, or jobs following high throughput computing [27]. By doing so, they can find the expected solution in a more cost-effective way, and perhaps discard unnecessary work to be processed.

Figure 1 presents the *submission-execution-analysis workflow* investigated in this scenario. In Step 1, the user submits to Copper a request to run an application with her SLA constraints, which can contain deadline, cost, or result quality restrictions. In this work we focus on deadline constraints. Others restrictions (e.g., cost or energy) can be mapped into the time it takes to execute jobs on the computing infrastructure. In a scenario where Copper is executed in the cloud, cost may become an important factor to be considered. Users may decide to change their execution strategies so as to meet cost constraints based on the cloud provider instance prices and charging model having job execution time and resource requirements as input. Another example, an energy utilization rate can be employed to define the maximum amount of energy to run the experiments depending on the execution time. The user also describes in the request the parameters and possible values to run her application, the job submission strategy (e.g., random search or range of parameter values), and the specification of the computing environment (e.g. number of processors and their configuration). In Step 2, Copper requests the resource management system to configure the environment and creates compute jobs corresponding to parameter-value pairs.

Once jobs are executing, Copper monitors the state of the running/completed/pending jobs to predict SLA violation (Step 3). If SLA is about to be violated, a new strategy is defined to run the pending jobs. If that is the case, a notification of SLA violation and a new strategy is sent to the user (Step 4) who, based on this information, can act on the strategy to execute her jobs (Step 5). Meanwhile, the analyst can also manually change job priorities based on her domain *expertise* and on her *trust* on the ongoing strategy (Step 6). As Copper exploits the analyst expertise and her feedback is not instantaneous, the job priority asynchronously changes to avoid unnecessary delays in triggering new jobs. This is particularly important as Copper is designed to manage long execution time jobs while users follow their submission-execution-analysis workflows.

The specific problems tackled in paper are therefore:

- How to suggest jobs for execution to the analyst and exploit her expertise to get faster problem resolution?
- How to predict SLA violations based on clustering strategies and pending/running/completed jobs?
- How to suggest to the analyst to change her strategy to avoid SLA violations based on intermediate results and her expertise?

## 3. ARCHITECTURE AND METHODS

The proposed solution comprises the Copper tool, a computing infrastructure, and a web client. Copper is a service, which can be executed in the cloud or on-premise, responsible for learning the user strategy, suggesting strategy changes in case of probable SLA violations, and proposing job executions to the analyst. A simple adapter should be created on the computational infrastructure to connect existing user applications and Copper. This component is specific to each application and allows Copper to trigger job execution by using REST commands. It is also responsible for sending intermediate results and the list of parameters with their possible values to Copper.

Figure 2 presents an overview of the Copper architecture. The *manager* orchestrates the Copper operation, reads the results, and stores them in a database as soon as jobs are executed. The *SLA evaluator* and *sampler* receive those intermediate results to update the selected DFO method state and *cluster* models. In order the produce new job candidates (§ 3.1), the sampler utilizes DFO strategies such as generalized greedy randomized adaptive search (GRASP), general pattern search (GPS), or particle swarm (PS) optimization [4, 10]. The analyst selects the optimization strategy according to its preference. In this paper, we implemented a GRASP variation that employs DOE to estimate the quality of a given sample. As the focus of this paper is related to the interaction between the analyst and Copper, the comparison between the developed GRASP version and other optimization approaches is beyond the scope of this paper.

For SLA violation prediction (§ 3.1), the *SLA evaluator* employs clustering methods to estimate the number of pending jobs according to the user strategy. This number is compared to the number of jobs that can be executed based on the remaining time/budget defined in the SLA contract. If the user has no resources to evaluate all pending jobs according to a given strategy, Copper sends a warning and suggests the user to change her approach.

Figure 3 shows Copper in the submission-execution-analysis workflow. Initially, the analyst selects the first jobs to be executed (initialEvaluation()). Those jobs can be selected at random by Copper or strategically selected by the analyst. Next, Copper submits the jobs to the computational infrastructure (submitJobs()). Once the first jobs are finished, Copper updates the internal DOE model (updateModel()) and presents the results to the analyst.

After the initial phase, in the Evaluation Loop the analyst submits the jobs and evaluates the results. The analyst initially receives a list of job suggestions (jobSuggestion()) from Copper and adapts the list by including or removing jobs according to her expertise (includeUserJobs()). Then, the jobs are submitted, the Copper model is updated, and the results are presented to the analyst (similar to the initial phase). Whenever the analyst considers the results reached the desired goal, the loop stops. If the ongoing strategy is predicted to lead to an SLA violation, the result of the jobSuggestion() method returns the violation warning and the new samples to be evaluated according to the analyst's predefined strategy (§ 3.2).

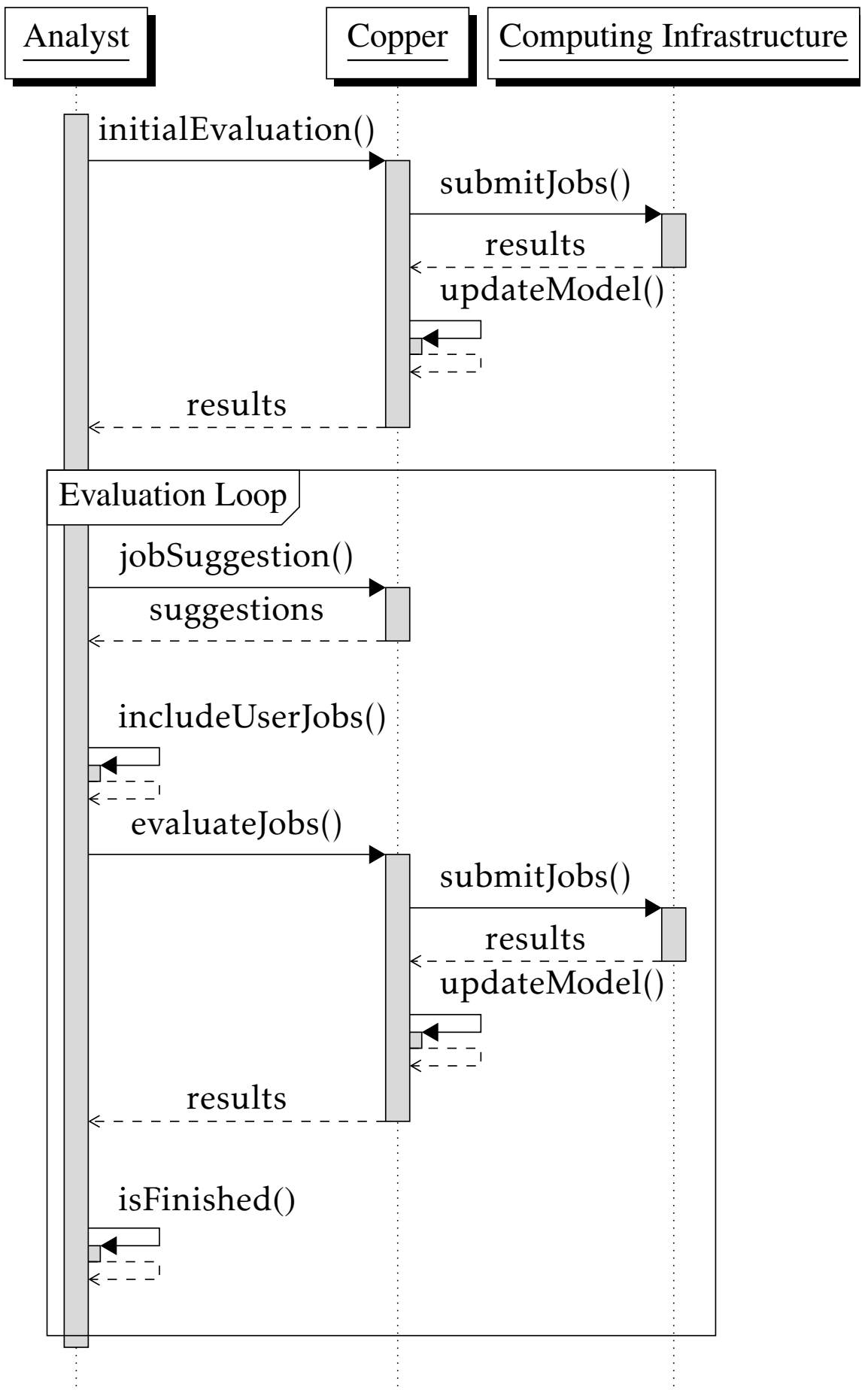

**Figure 3: Copper in the submission-execution-analysis workflow.**

**Algorithm 1:** *sampling* - proposed sampling method.

**Input** : $M$, $P$, $\beta$, $n$
**Output**: *samples*
1 solution ← construction($M$, $\beta$ , $P$);
2 samples ← localSearch($M$, $P$, $n$, *solution*);
3 return samples;

**Algorithm 2:** *construction* - algorithm for creating a solution candidate.

**Input** : $M$, $P$
**Output**: *solution*
1 solution ← ∅;
2 **foreach** *p of P* **do**
3 | $RCL \leftarrow \{e \in \text{listValues}(p, M) \mid q(e) \geq q^{min} + \beta(q^{max} - q^{min})\}$;
4 | solution[indexOf($p$)] ← getRandomElement($RCL$);
5 **end**
6 return solution;

**Algorithm 3:** *localSearch* - find neighbors of a given solution.

**Input** : $M$, $P$, $n$, *solution*
**Output**: *solutions*
1 solutions ← ∅;
2 **for** $i \leftarrow 1$ **to** $n$ **do**
3 | solutions[i] ← getNeighbor($M$, $P$, *solution*);
4 **end**
5 return solutions;

## 3.1 Default Sampling Method - GRASP

Combinatorial optimization corresponds to a class of problems that consist of finding an optimal solution from a finite set of objects [31]. Generally, brute-force approaches cannot be applied to solve those problems due to time or cost constraints. Therefore, analysts or computer programs should employ heuristics to select candidate solutions that will probably represent good (or optimal) solutions. This section presents how Copper generates evaluation *samples*[1] to the analyst using GRASP algorithm. This method is based on the original GRASP algorithm [10] but instead of automatically evaluating the produced samples, the analyst decides whether the samples will be assessed or not. This sampling approach corresponds to the default method to suggest evaluation of the search space samples to the user. Our approach is not strictly tied to this optimization method and other techniques can be adopted to generate user samples (e.g., GPS or PS).

The GRASP sampling method is presented in Algorithm 1, which comprises two phases: construction and local search. The construction phase builds a feasible solution for each algorithm execution (Line 1). Next, the local search generates the neighbors of the previous solution (Line 2). Finally, the set of results is returned (including the initial solution). The algorithm inputs are the following: the DOE model ($M$), the list of parameters ($P$), the greediness level ($\beta$), and the number of generated samples ($S$). The greediness level ($\beta$) is a real value contained in $[0, 1]$ and represents how random is the selection of parameter values according to their estimated quality. If the greediness level is close to one, only high-quality parameter values can be selected. On the other hand, $\beta$ values close to zero allow the selection of a broad range of values including low-quality values.

The construction phase is presented in Algorithm 2 and finds a candidate based on the greedy heuristic. Initially, an empty solution is created (Line 1) to return the list of values for each parameter of the result solution. For each parameter $p$ (Line 2), a reduced candidate list (RCL) is created and this list is randomly sampled to get a value that composes the final solution.

We use the full factorial DOE method [21] to evaluate the quality of each parameter value. $q^{min}$ and $q^{max}$ are the minimum and maximum quality values of a given parameter value estimated by the DOE method. The RCL is formed by all feasible parameter values that can be used to construct the partial solution. In order to insert a parameter value in the RCL (Line 3) its quality level ($q$) should be higher or equal than a threshold $q^{min} + \beta(q^{max} - q^{min})$. Observe that if $\beta \sim 1$, the quality threshold approximates to $q^{max}$ and only the best quality results will be selected (pure greedy approach). On the other hand, if $\beta \sim 0$, the quality threshold is close to $q^{min}$ and most of the parameter values can be used to compose the final solution (i.e., random approach).

Whenever part of the search space has enough evaluated samples to apply the full factorial, the DOE model is updated. Next, a component of RCL is randomly sampled to compose the final solution (Line 4), and finally the solution is returned (Line 6).

The last phase of the sampling algorithm corresponds to the local search (Algorithm 3). In this phase, $n$ samples close to the candidate solution found in the previous phase are selected. The getNeighbor() function is employed to get a solution candidate neighbor by using euclidean distance between samples.

## 3.2 SLA violation prediction method

The method to predict SLA violations is based on clustering techniques. We identify the clusters of the search space that the analyst is interested in and compare the number of pending jobs in these clusters to the number of jobs that can be executed according to the SLA contract. In this work, we use Density-Based Spatial Clustering of Applications with Noise (DBSCAN) [14] method to identify the analyst strategy. This clustering method is attractive as the analyst does not need to inform the number of clusters before evaluation [31]. The DBSCAN algorithm groups samples with many neighbors determining high density clusters. Samples that lie in low-density clusters are defined as outliers and may not be classified in any cluster. Here this algorithm is adopted to determine previously evaluated clusters of the search space with good solutions and suggests these clusters in case of possible SLA violations. The user defines how the pending jobs will be distributed over the identified clusters.

Figure 4 presents the basic idea of the proposed solution. Suppose the search space is composed of two parameters and each particular combination $(v1, v2)$ represents a job that will be submitted to the computational infrastructure. $v1$ and $v2$ are possible values for parameters 1 and 2 respectively. As long as the analyst submits new jobs, the clus-tering approach identifies the clusters of the search space that provide good results. The quality of job results can be previously defined by using a threshold value or the analyst can qualify each job result (e.g., good or bad) whenever it becomes available.

[1]Samples and jobs represent an instance of the application with a set of parameter values. These terms are used in the paper interchangeably.

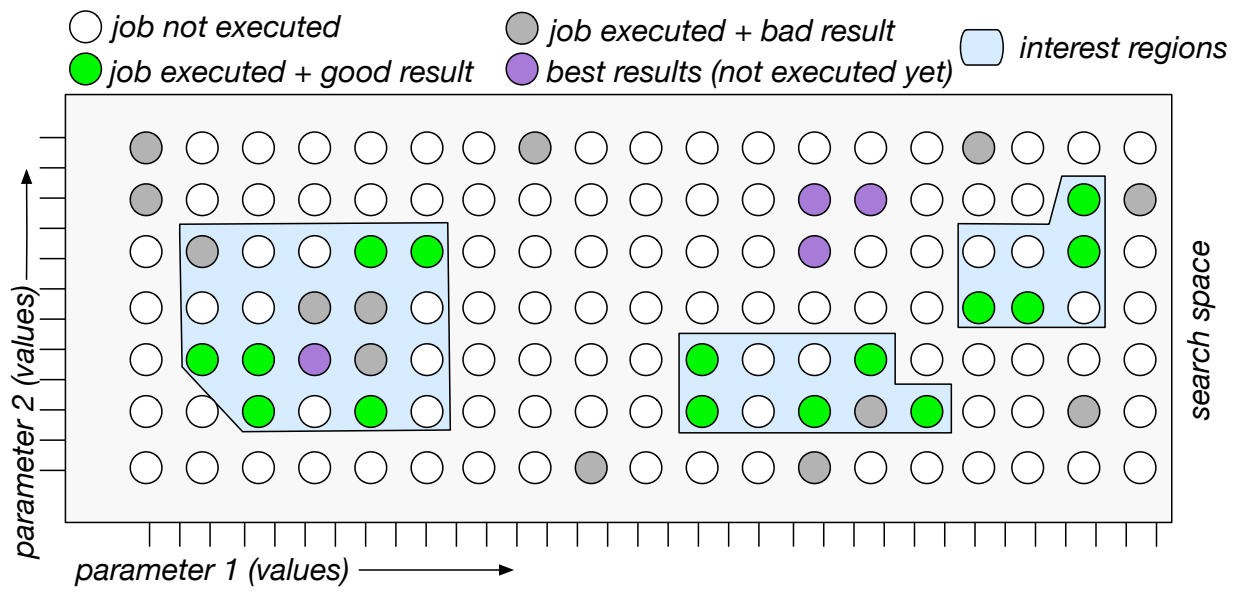

**Figure 4: Example of search space with a set of interest clusters.**

In this example, the analyst is searching the best result in three clusters of the search space. The clusters are identified by the clustering algorithm and the number of pending jobs is counted. If this number is higher than the remaining jobs defined by the SLA the user should change her strategy. In this case, the number of pending jobs is 15 (including the best result). If the analyst can run 15 jobs, at most, jobs due to SLA restrictions, Copper advises the analyst to change her strategy.

The strategy can be defined considering *exploration* and *exploitation* aspects. We define exploitation level ($e_x$) as the proportion of pending jobs that will be executed considering samples inside the detected clusters. For instance, if the exploitation level is 1.0 all the pending jobs will be executed taking the parameters contained in the clusters of interest. If the exploitation level is 0, pending jobs in not explored clusters (outside the clusters) will be given higher priority.

Considering all possible pending jobs in all clusters (*i.e.* $e_x \times$#pending jobs), the user defines which ones of these jobs will execute for each cluster. The default approach utilizes weighted average according the mean quality of evaluated jobs inside the clusters. For jobs outside the clusters (*i.e.* $(1 - e_x) \times$ #pending jobs), the default GRASP strategy remains the same (§ 3.1).

# 4. EVALUATION

In this section, we evaluate the Copper effectiveness in helping users run optimization and computer simulations under SLA constraints. We first describe three applications served as use cases and metrics analyzed to understand the benefits of the assistant. In particular, we are interested in analyzing three major aspects considering submission-execution-analysis workflows: (i) the interaction of Copper with analysts of different expertise levels; (ii) how the analyst confidence on Copper affects the evaluation of user applications; (iii) the impact of Copper SLA violation predictions on user experiments.

## 4.1 Experiment Setup

This section presents an overview of the setup to evaluate Copper.

### 4.1.1 *Applications*

Here is a short description of the applications we used in the evaluation with an overview of their input parameters and output users are interested in.

**1. ifm:** The Integrated Flood Model (IFM) [28] is a hydrological model aimed at providing high resolution flood forecasts. IFM contains a soil and an overland routing model where the soil model estimates the surface-runoff based on incoming precipitation, soil, and land use properties. When IFM is deployed to calculate flooding predictions, it requires a calibration which consists in running several simulations that match data collected by real sensors. To evaluate IFM, we varied the parameters related to incoming precipitation and soil properties with the goal of minimizing the difference between simulated and real data captured by sensors.

**2. schedsim:** Scheduler Simulator (SchedSim) is a simulator to assist in the creation of policies to manage High Performance Computing clusters. It accepts a variety of parameters including number of processors in the cluster, partitions, and scheduling algorithms. Tuning the scheduler and cluster properties to meet client business goals is not a trivial task and several scenarios must be executed to achieve that. For the evaluation, we had the following input: (i) a workload containing historical data of jobs submitted to a cluster; (ii) a fixed number of cluster processors; and (iii) two variable partitions. We wanted to know what was the partition sizes and which jobs (requested time and allocated processors) should go to such partitions having the following goals: (i) minimize the overall job response time and (ii) small jobs (under 1h requested time) should wait no more than 3 hours in the cluster waiting queue.

**3. mazerunner:** Maze Runner is an in-house created game in which simulated runner has to find the way out of a maze. The objective of the game is to find a runner configuration (e.g., velocity, movement strategy) that leads to a shorter time to find the maze exit. There are enemies in the maze, and the runner needs to escape from them. Additionally, the runner can slip on obstacles and stay vulnerable for some time depending on her velocity. This game is simple and intuitive and was created to demonstrate concepts of the proposed assistant for a broader audience.

The search spaces related to the applications used in this work are illustrated in Figure 5. Multiple parameter combinations are represented in a single axis and each square corresponds to a parameter combination. The quality of a parameter configuration is represented as a given color (dark colors represent better results). For instance, the top-right square of Figure 5c presents the (*V1_1, V2_1, V3_2, V4_3*) parameter configuration of maze runner. In this work, we assume the output of parametric applications as black box functions, therefore we will not enter in details about the meaning of parameters and their values.

By observing the search spaces of the three applications, SchedSim presents a well behaved variation of the output when the parameters values change. A similar behavior happens to the IFM search space. However, for higher values of *P1* the output presents no variation. This 'flat' search space region reduces the Copper capacity to predict good parameter values. Regarding the Maze Runner search space, configurations whose parameter *P1* = *V1_3* provide better results. However, the output variation does not have the same gradual characteristics as in the other applications.

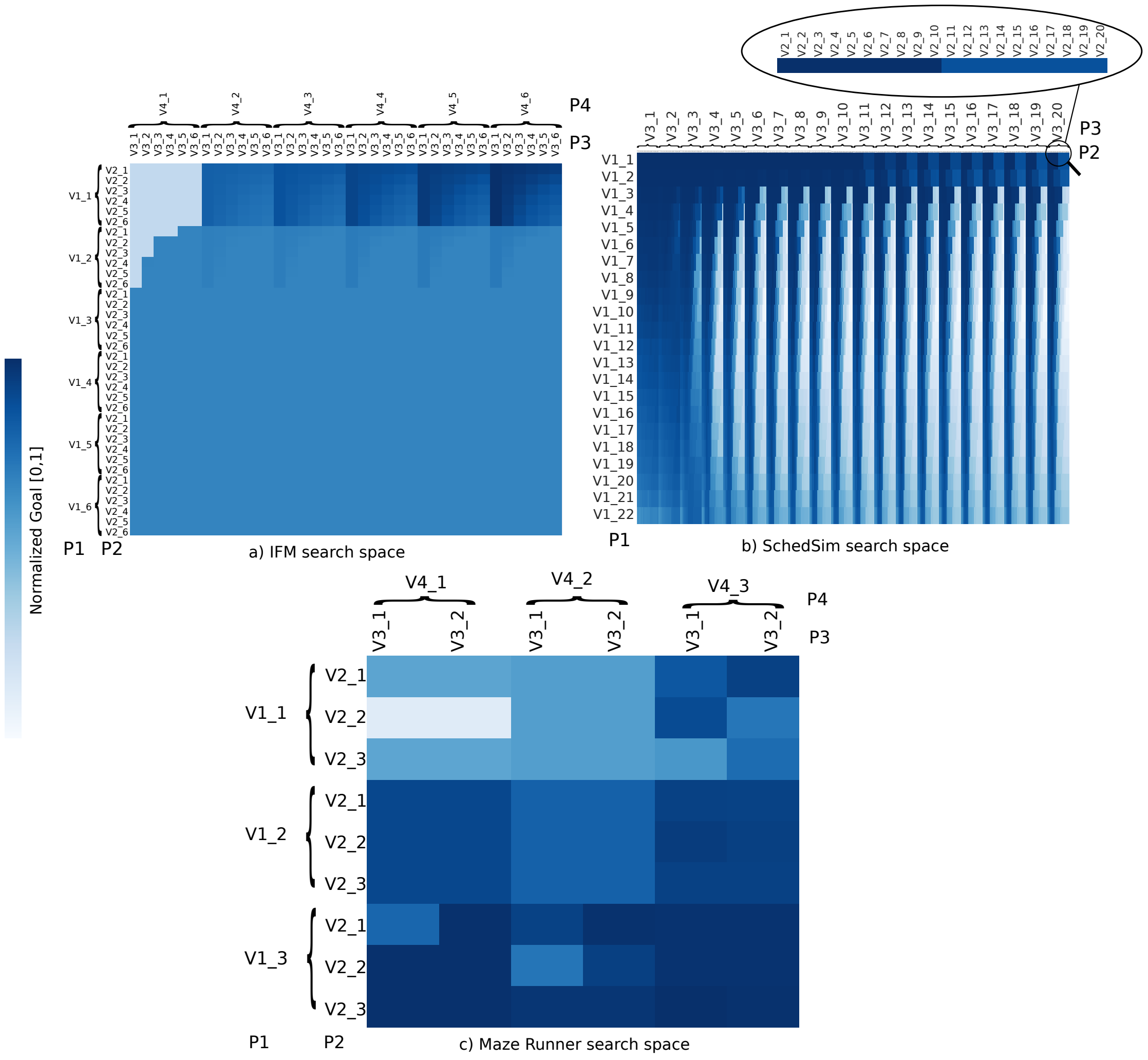

**Figure 5: Search spaces for IFM, SchedSim and Maze Runner applications. Each application has $n$ parameters ($P1 \cdots Pn$) and each parameter $Pj$ can assume $m$ values ($Vj_1, \cdots, Vj_m$). The search spaces sizes for IFM, SchedSim, and Maze Runner are $6 \times 6 \times 6 \times 6 = 1296$, $22 \times 20 \times 20 = 8800$, and $3 \times 3 \times 2 \times 3 = 54$, respectively.**

#### *4.1.2 Experiments Description*

We consider software agents to simulate the human behavior [19] and divide the experiments into two main groups. First, the agents evaluate the tool with no SLA requirements (§ 4.2). In this case, the assessment aims at finding how many jobs are necessary to find an optimal solution. The software agent only interacts with Copper to obtain suggestions of jobs to be executed and no SLA-aware assessment is performed.

The second evaluation is performed to find the best result considering a limited number of jobs (§ 4.3). Therefore, agents interact with Copper to obtain strategy suggestions if the ongoing experiments tend to extrapolate the number of jobs established in the SLA.

## 4.2 Finding an Optimal Solution

In this evaluation, we assess the number of jobs that must be executed to find an optimal solution. The agents have insights about which parameter values may result in optimal solutions. Therefore, we divided the agents into three categories based on the insight quality level. The first group of agents are related to users that have a previous intuition about good parameter values, and this intuition may lead to bad solutions (*bad insights*). The second group contains agents that have previous ideas about the parameter values and these values result in proper solutions (*good insights*). The last group of agents represents users with no insight about the clusters of search space that may contain the optimal solution (*no insights*).

For each agent insight level, we vary the user confidence level on the job suggestions that come from Copper and the capacity of an agent to learn from intermediate results. The confidence level determines the percentage of jobs that are submitted taking into account Copper suggestions. For instance, if the user submits 100 jobs and the confidence level is 40%, then 40 jobs are submitted based on Copper recommendation and the rest is triggered based on the agent intuition. This intuition corresponds to a subset of search space in which the analyst believes the optimal result is contained.

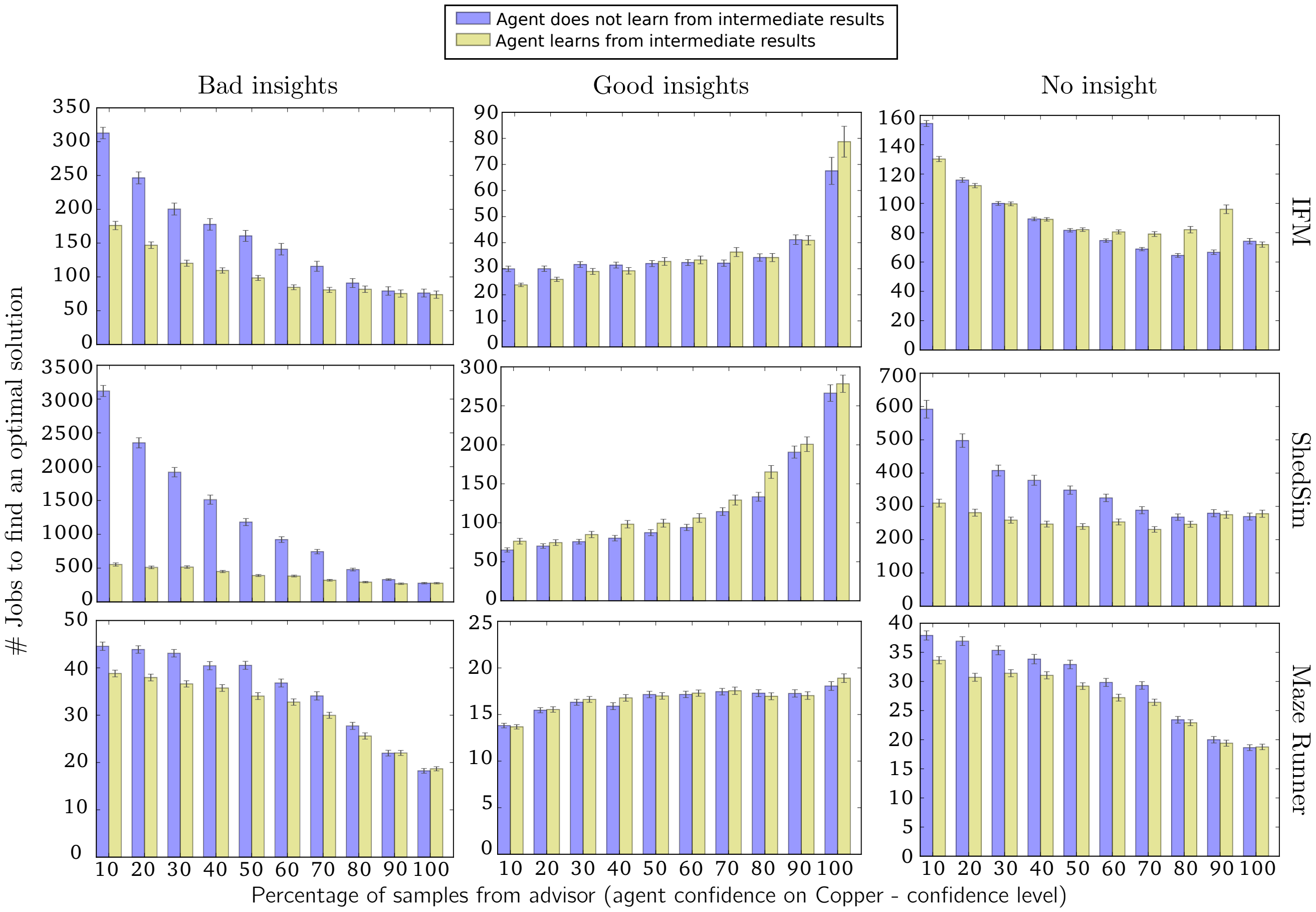

**Figure 6: # Jobs to find an optimal solution as a function of agent confidence level for all applications. 100% confidence level experiments correspond to the proposed GRASP solution without analyst intervention.**

We evaluate agents that learn or not from intermediate results. When the agent learns from intermediate results, it increases the probability of selecting a given parameter value if this value leads to good intermediate results and decreases the likelihood of choosing a given value if this parameter value leads to bad results. As the focus of this experiment is to present how the framework takes benefits from users recommendations, we do not explore in details here how the agents learn from intermediate results.

Figure 6 shows the percentage of jobs that are submitted according to Copper recommendation (confidence level) for all applications as a function of the number of jobs that must be submitted to find an optimal solution. As the proposed sampling algorithm is stochastic, multiple experiments were performed to find a 95% confidence interval of the mean number of jobs to find an optimal result. The bootstrap method [8] was adopted to find the confidence interval and the sample size for each result is 500—this number is high and reduces the size of the confidence interval. The bar graphs show the mean value and the confidence interval for each experiment.

The first, second, and third rows of Figure 6 are related respectively to IFM, SchedSim, and Maze Runner results. The three columns of the same figure, shows the experiments for agents with bad, good and no previous insights for these applications. For agents with bad insights, as the confidence level on recommendations made by Copper increase, the number of jobs to find an optimal result reduces. Then, users with bad insights may have benefits when using Copper. For Agents that do not learn from intermediate results, SchedSim results present the most significant difference between relying on Copper to provide the jobs parameters or not. Increasing the confidence in suggestions made by Copper also reduces the number of jobs to find the optimal configuration in IFM and Maze Runner. However, this difference is less significant than the previous one. As the SchedSim search space has interesting characteristics about the objective function growing related to the input parameters, the proposed sampling algorithm takes less steps to find the optimal parameter configuration.

For bad insight agents that learn from intermediate results, there is a significant difference between learning from intermediate results or not. For example, bad insight agents for SchedSim with 10% confidence level presents a big difference between learning from intermediate results or not. This behavior can be explained by the interesting characteristics of ShedSim search space. Figure 5 shows that low values of SchedSim parameters *P1*, *P2*, and *P3* lead to good results. The agents learn that these values lead to good results and increase the probability of choosing combinations of these values.

For agents with good insights, the number of jobs to find the optimal solution increases with the agent confidence level. This result is the opposite of the previous one and indicates that users with good insights may have a significant reduction of steps to find an optimal solution. Agents with good insights about the behavior of SchedSim and Maze Runner present a regular growth in the number of jobs to find the optimal solution when the confidence level on Copper increases. However, IFM does not present this characteristic as there is an abrupt increase from 90 to 100 percent of confidence level. Whenever 100% confidence level is employed, the agent only takes Copper suggestions to submit new jobs. In this case, Copper needs more jobs to select parameters outside the flat area of IFM search space. As 90% confidence level agents submit 10% of jobs based on user insight. The DOE model is updated considering the non-flat areas (agent good insight) and speeds-up the optimization process. For agents with good insights, there is no expressive difference between learning from intermediate results or not. As the number of jobs to find the optimal solution is reduced, the number of samples is not sufficient to update the user preference and significantly affect the result.

For agents with no insights, the agent's efficiency is impacted by the confidence level. Whenever the confidence level on Copper increases the results get better. However, this impact is less notable to no insight agents when compared to the results of bad insight agents. Similarly to the bad insight agents, there is a difference between learning from intermediate results.

## 4.3 Restricted Experiments

This set of experiments utilizes the same agents to find the best effort solution but with limited number of jobs that can be executed due to SLA constraints. We assume that all jobs take the same time to execute, and the SLA constraints correspond only to time requirements. Other conditions such as costs or energy consumption can be used by mapping these requirements into the number of submitted jobs. The quality of a result is given by the ratio of the result found and the optimal result.

For each experiment set, we changed the maximum number of jobs that can be submitted and the exploitation level. The exploitation level is defined as the proportion of pending jobs that will be submitted considering samples inside the clusters defined in § 3.2. As the output of an experiment is not deterministic, each bar in the chart presents the mean value and the 95% confidence related to 500 runs for each maximum jobs, and exploitation level pair. In these experiments, the proposed clustering method is applied when the number of submitted jobs reaches 70% of the maximum number of jobs and the agent accepts the Copper recommendation for all experiments. The restricted experiment results for IFM, SchedSim, and Maze Runner are respectively presented in Figures 7, 8, and 9.

The IFM experiment set examines the best effort scenario with a limited number of jobs. In this case, instead of preferring exploration or exploration of sampling space, it is preferable to adopt a mixed strategy to find good parameter values. For instance, for 50 jobs the experiment with exploitation level of 33% presents better quality when compared to the others. Similar to the other experiment sets, the variation of results quality is higher when the number of jobs is reduced (in this case 50 jobs). Based on these results we can conclude that for this experiment it is preferable to adopt a mixed strategy to evaluate the search space. Therefore, experienced analysts can take advantage of combined search strategies of exploitation/exploration to achieve better results.

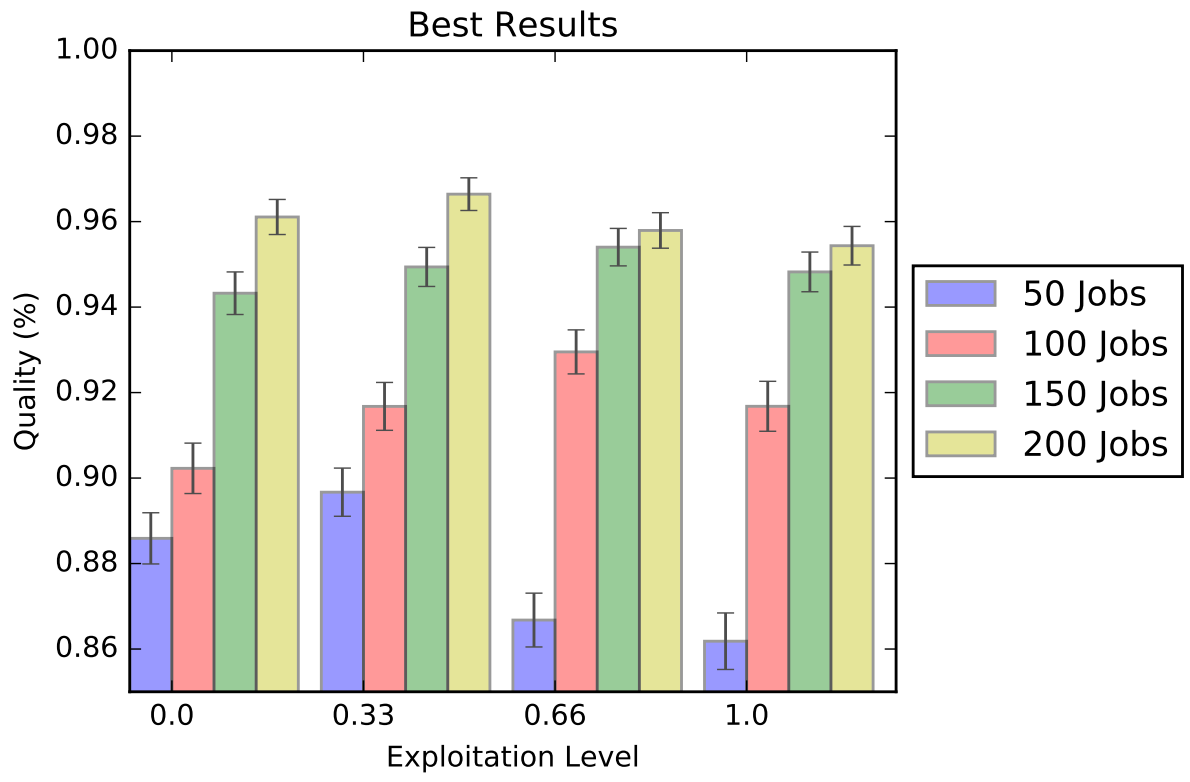

**Figure 7: Best results for experiments with SLA restrictions - IFM.**

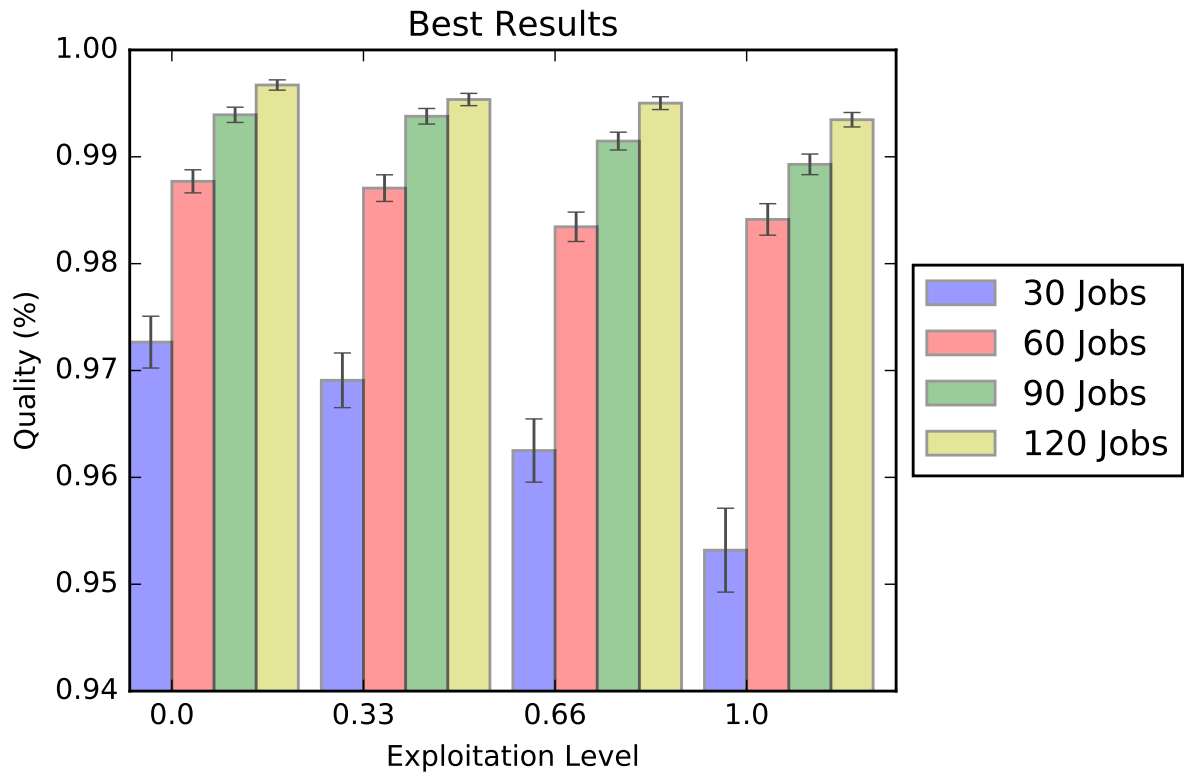

**Figure 8: Best results for experiments with SLA restrictions - SchedSim.**

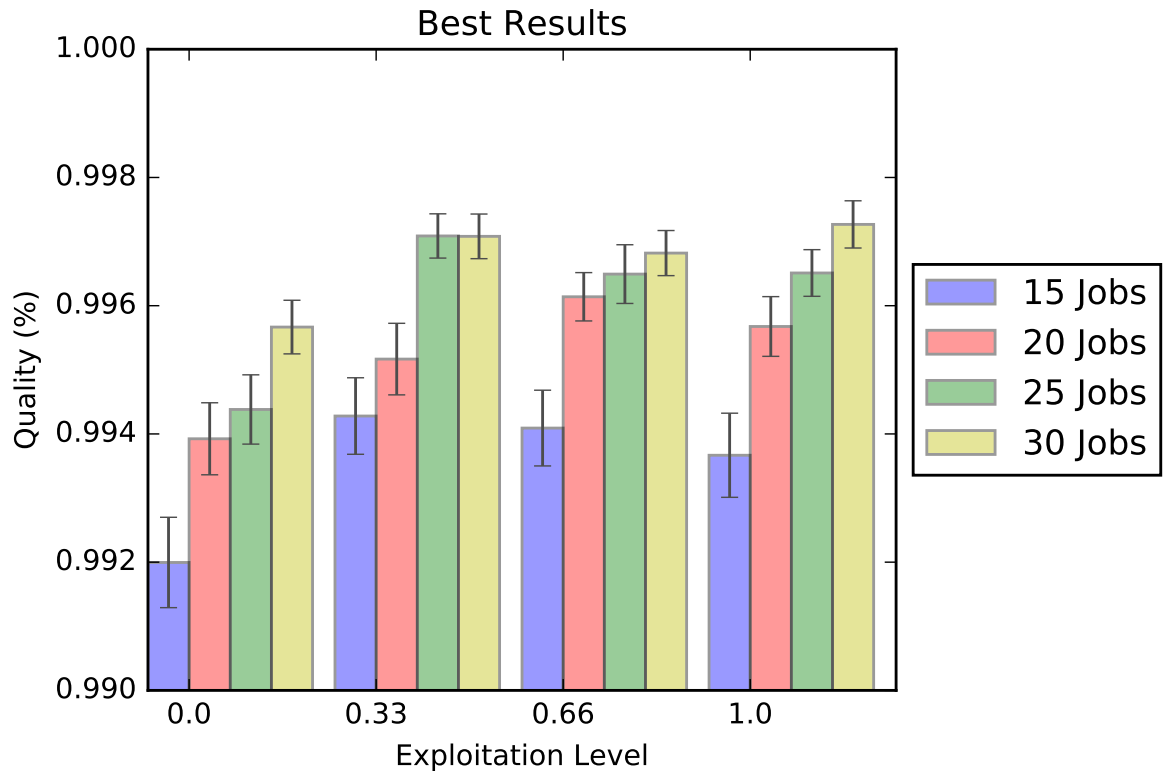

**Figure 9: Best results for experiments with SLA restrictions - Maze Runner.**

For the SchedSim results (Figure 8), it is possible to observe that for a reduced number of jobs (i.e., 30) the exploitation level has a significant influence on the output quality. The impact of exploitation level decreases as the number of submitted jobs increase because the probability of finding better results are higher if more jobs are executed. For this experiment set, it is preferable to explore the search space instead of concentrating the submitted jobs in the cluster areas. This result suggests that the best results are spread through the solution search space.

Figure 9 presents the results of experiments with a restricted number of jobs related to Maze Runner. Different from the previous application, whenever the exploitation level is higher, the results have better quality. On the other hand, when the SLA constraint is relaxed (more jobs are allowed) the impact of exploitation levels becomes less important. We can conclude that the best results are grouped in the solution search space, then strategies that exploit intermediate results can be adopted to get better results.

# 5. CONCLUSIONS

Several areas in science and engineering rely on many executions of a software system with different values for each supported parameter. These executions are responsible for evaluating test scenarios and calibrating models. Normally, users follow a workflow where they setup a set of experiments, generate intermediate results, analyze them, and reprioritize jobs for the next batch. Reprioritization is a key step as executing all possible values for all supported parameters is usually not feasible, especially under cost and deadline constraints determine by strict SLAs.

This work introduced a tool to help users in executing their software systems by exploring their domain expertise and design of experiment techniques. By learning user strategies on sweeping parameters, our tool is able to detect if the ongoing strategy is able to meet SLA requirements and suggest how jobs can be reprioritized in case an SLA violation is predicted to happen while users run their submission-execution-analysis workflows. Our main lessons while developing the tool and evaluating with three applications are: (i) it is important to facilitate user interaction with an automated design of experiment technique as users may bring domain expertise that can reach desired results faster; (ii) as the desired solution may be found by mixing exploration and exploitation strategies (e.g. breadth-first search and depth-first search), users can benefit from a tool that can identify if the ongoing strategy is correct and can meet SLA constraints; (iii) evaluations with a restricted number of experiments are highly impacted by Copper suggestions. Then, the proposed tool and methods are especially suitable for evaluation of parametric applications with tight deadlines.

## Acknowledgements

We would like to thank Miguel Paredes Quinones and the anonymous reviewers for their valuable comments. This work has been partially supported by FINEP/MCTI under grant no. 03.14.0062.00.